\documentclass[twocolumn,showpacs,prl]{revtex4}

\usepackage[dvips]{graphicx}% Include figure files
\usepackage{dcolumn}% Align table columns on decimal point
\usepackage{bm}% bold math

\begin{document}

\title{Ettingshausen Effect around Landau Level Filling Factor \boldmath{$\nu=3$}\\
Studied by Dynamic Nuclear Polarization}

\author{Yosuke Komori}

\author{Satoru Sakuma}

\author{Tohru Okamoto}

\affiliation{Department of Physics, University of Tokyo, 7-3-1 Hongo, Bunkyo-ku, Tokyo 113-0033, Japan}

\date{13 March 2007}

\begin{abstract}
A spin current perpendicular to the electric current is investigated around a Landau level filling factor $\nu=3$ in a GaAs/AlGaAs two-dimensional electron system.
Measurements of dynamic nuclear polarization in the vicinity of the edge of a specially designed Hall bar sample indicate that the direction of the spin current with respect to the Hall electric field reverses its polarity at $\nu=3$, where the dissipative current carried by holes in the spin up Landau level is replaced with that by electrons in the spin down Landau level.
\end{abstract}
\pacs{73.43.-f, 72.25.Pn, 73.50.Jt, 76.60.-k}

\maketitle

A two-dimensional electron system (2DES) at low temperatures and 
in a strong magnetic field shows the quantum Hall (QH) effect,
in which the longitudinal resistance vanishes and the Hall resistance is quantized as $R_{H}=h/ie^2$ for an integer $i$ \cite{Kawaji1980,Klitzing1980}.
While the electric current does not involve energy dissipation in the QH state, it can cause heat flow in the transition region between QH states and in the current-induced breakdown regime.
Heat flow parallel to the electric current has been studied in the current-induced breakdown regime by measuring the longitudinal resistance with a set of voltage probes along the current channel \cite{Komiyama1996,Kaya1998,Morita2002}.
On the other hand, Akera predicted that the heat flow in the current-induced breakdown regime can have a component perpendicular to the electric current \cite{Akera2002}.
In general, this phenomenon is known as the Ettingshausen effect.
His calculation shows that the heat flow across the current channel causes an increase (decrease) of the electron temperature $T_e$ in the vicinity of one edge (the other edge) while $T_e$ is approximately uniform in the middle of the current channel.
Furthermore a recent calculation for the transition region shows that the sign of the electron temperature gradient perpendicular to the electric current exhibits quantum oscillations as a function of the position of the chemical potential $\mu$ with respect to the Landau levels (LLs) \cite{Akera2005}.
Evidence of the Ettingshausen effect is observed in the current-induced breakdown regime at Landau level filling factor $\nu=2$ by Komori and Okamoto, who used micro-Hall bars attached to both edges of the current channel as electron temperature indicators \cite{Komori2005}.
However, they were not able to extend their work to the transition region owing to the strong current-dependence of the background at $\nu \neq {\rm integer}$.

In a strong magnetic field, the spin degeneracy is lifted due to the Zeeman energy and the many body effect.
When $\mu$ lies between spin-split Landau levels, electron spin polarization strongly depends on $T_e$ and the heat flow is dominated by the spin current.
One of the advantages of using the spin degree of freedom in the study of heat flow is that the flip of electron spin can be memorized by dynamic nuclear polarization (DNP) owing to the contact hyperfine interaction \cite{Dobers1988,Berg1990,Wald1994,Dixon1997,Kronmuller1998,Kronmuller1999,Eom2000,Kraus2002},
\begin{eqnarray}
A {\mathbf I} \cdot {\mathbf S} = \frac{A}{2} [ I_+ S_- + I_- S_+] +AI_zS_z,
\end{eqnarray}
where $A$ ($>0$) is the hyperfine constant, and ${\mathbf I}$ and ${\mathbf S}$ are the nuclear spin and electron spin, respectively.
%XXXXX second term
The first term causes the electron-nucleus flip-flop process and the second term changes the electron Zeeman energy effectively.
%XXXXX ref9
In Ref.~[\onlinecite{Dobers1988}], DNP about 8~\% of the maximum polarization, which produces the effective magnetic field of 0.43~T, was achieved from the Overhauser shift of the electron spin resonance line.
Nuclear spins in semiconductors have attracted great attention due to the possible application for quantum information technology \cite{Kane1998,Bennett2000}.
Electrical local manipulation of DNP  has been intensively studied by several groups \cite{Smet2002,Hashimoto2002,Machida2002,Wurtz2005}.
Very recently, two of the present authors have demonstrated that DNP is induced by an electron temperature change at $\nu=3$ in a narrow channel sample where the width varies stepwise along the current flow \cite{Komori2007}.

In this work, we study the spin current perpendicular to the electric current around $\nu=3$ in a GaAs/AlGaAs 2DES using DNP.
Consider the spin current running along the $x$ axis perpendicular to the electric current carrying channel parallel to the $y$ direction.
At both edges, it should be terminated by spin flips, which induce DNP.
If the $x$ component of the spin current is positive, electron spin flips from up to down occur predominantly and positive DNP is induced in the vicinity of the right (large-$x$) side edge.
From the sign of DNP, which is different for both edges, we can deduce the polarity of the spin current.
The width of the DNP region determined by the characteristic spin relaxation length is expected to be of the order of 1~$\mu$m in the system studied \cite{Komori2007}.
%XXXX refs,,,
In Refs.~[\onlinecite{Wald1994,Dixon1997,Machida2002}], the DNP in the vicinity of the edge was successfully detected through a change in the tunneling rate between spin resolved edge channels in specific devices with the front gates.
In the present work, we used the edge-bulk coupling and the detection of DNP was performed at a fixed value of $\nu=3.2$.
Long longitudinal relaxation time $T_1$ of DNP allows us to change $\nu$ from an arbitrary value, at which DNP is induced by a large electric current, to 3.2 by sweeping the magnetic field.
As illustrated in Fig.~1, at $\nu=3.2$, the spin down LL with the orbital index $n=1$ [LL($\downarrow, 1$)] is partially filled and makes the resistive bulk region while the spin up LL with $n=1$ [LL($\uparrow, 1$)] forms edge channels.
In this situation, the backscattering between opposite edge channels, which is proportional to the longitudinal resistance, is strongly affected by the degree of equilibration between the edge channels and the bulk region \cite{Kane1987,Komiyama1993,Hirakawa2001}.
Positive (negative) DNP effectively decreases (increases) the electron Zeeman energy via the second term in Eq.~(1) \cite{Komori2007} and is expected to enhance (reduce) the edge-bulk coupling.
Since the effect of the right side DNP on the longitudinal resistance is canceled by that of the left side DNP if we use a symmetric configuration, we designed an asymmetric Hall bar where only the right side DNP affects the resistance change.
The observed longitudinal resistance change due to DNP indicates that the direction of the spin current with respect to the Hall electric field depends on $\nu$.
\begin{figure}[t!]
\includegraphics[width=5cm]{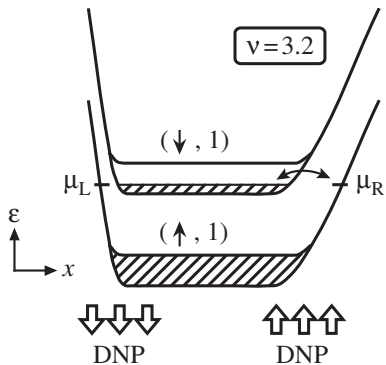}
\caption{
Spin-split LLs at $\nu=3.2$. A temperature broadening of the Fermi distribution function and LLs with $n=0$ are not drawn for clarity. Here we use an asymmetric sample geometry to detect DNP in the vicinity of the right edge. 
}%XXXXX
\end{figure}
\begin{figure}
\includegraphics[width=6cm]{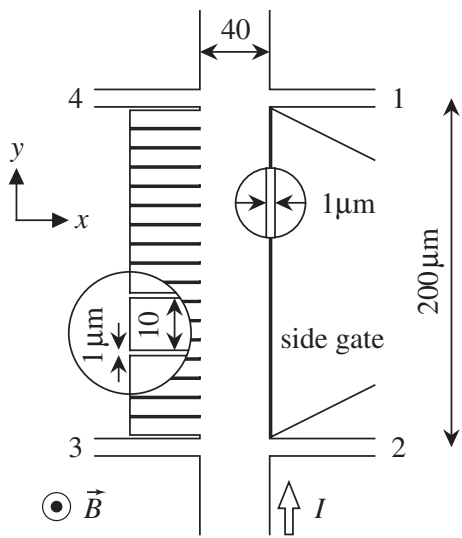}
\caption{
Schematic view of the central part of the sample.
The side gate is formed by chemical etching and isolated from the right edge of the Hall bar by a gap of $1~\mu$m.
}%XXXXX
\end{figure}

The sample was fabricated from a GaAs/Al${}_{0.26}$Ga${}_{0.74}$As heterostructure having an electron density of $4.9 \times 10^{15}~{\rm m}^{-2}$ and a mobility of $70~{\rm m}^2 / {\rm V~s}$ after brief illumination with a red light-emitting diode at 4.2~K.
Figure~2 shows the geometry of the central part of the sample.
In order to avoid the effect of hot spots, $600~\mu {\rm m}$ wide current electrodes are separated from the central part by $1200~\mu {\rm m}$ \cite{Kawaji1994}.
The confining potential profile of the right edge can be controlled by applying negative voltage to the side gate region isolated by a gap of $1~\mu$m.
The meander-shaped edge on the left side is designed so as to obtain the equilibration between the edge channel and the bulk region.
The longitudinal resistance $R_{yy}$ was measured by monitoring the voltage between contacts 1 and 2.
A standard low frequency lock-in technique was used.
The results were similar when contacts 3 and 4 were used.
All the measurements were performed at 1.6~K in a liquid helium cryostat.
The temperature was precisely controlled by monitoring the vapor pressure.

In Fig.~3, the $B$ dependence of $R_{yy}$ at a small ac electric current $I_{\rm ac}=0.1~\mu {\rm A}$ is shown for different side gate voltage $V_{\rm SG}$.
By applying negative $V_{\rm SG}$, although the $B=0$ resistance slightly (by 2.6~\%) increases from $V_{\rm SG}=0~{\rm V}$ to $-1~{\rm V}$ due to the reduction of the effective width, $R_{yy}$ drastically decreases in the lower-$B$ side of the QH minima.
This is attributed to the suppression of the edge-bulk coupling due to a gradual slope of the confining potential induced by the side gate.
The bulk magnetoresistance recovers when the electric current becomes large \cite{Kane1987,Komiyama1993,Hirakawa2001}.
It was confirmed that the $V_{\rm SG}$ dependence of $R_{yy}$ is very small for $3<\nu<4$ when $I_{\rm ac}=10~\mu {\rm A}$ is driven.
For the detection of DNP, we optimized the experimental conditions so as to obtain a large resistance change and used $V_{\rm SG}=-1.0~{\rm V}$ with $I_{\rm ac}=0.1~\mu$A and $\nu=3.2$.
\begin{figure}
\includegraphics[width=7cm]{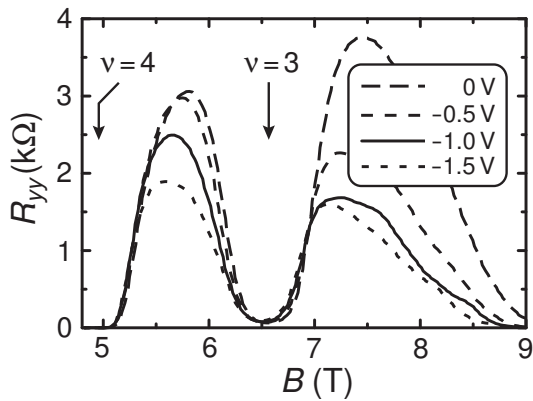}
\caption{
$R_{yy}$ in a perpendicular magnetic field, taken at $I_{\rm ac}=0.1~\mu {\rm A}$ and $T=1.6~{\rm K}$ for different $V_{\rm SG}$.
The arrows indicate minima corresponding to integer quantum Hall states.
}%XXXXX
\end{figure}

Figure 4 shows typical time evolution of $R_{yy}$ after applying a large dc current $I_{\rm dc}=+10~\mu {\rm A}$ for 10~min at $\nu=2.8$ or 3.4.
\begin{figure}
\includegraphics[width=7cm]{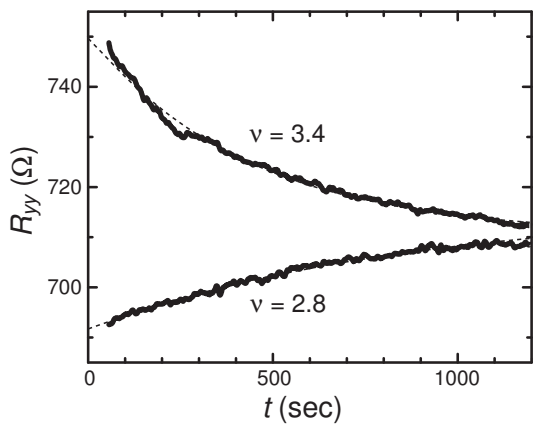}
\caption{
Time evolution of $R_{yy}$ at $\nu=3.2$ and $T=1.6~{\rm K}$ after applying a large dc current $I_{\rm dc}=+10~\mu {\rm A}$ for 10~min at $\nu=2.8$ or 3.4.
Magnetic field sweeping was done within 1~min after $t=0$.
Single exponential fits indicated by the dotted lines are used to estimate $R_{yy}$ at $t=0$.
}%XXXXX
\end{figure}
Deviation of $R_{yy}$ from the equilibrium value is clearly seen while its sign depends on $\nu$.
As discussed above, positive (negative) resistance change corresponds to the enhancement (reduction) of the edge-bulk coupling caused by positive (negative) DNP in the vicinity of the right edge and indicates positive (negative) $x$ component of the spin current.
The relaxation time $T_1$ is found to be 300-800~sec which is of the same order as that obtained at 1.6~K in Ref.~[\onlinecite{Komori2007}].
The nuclear origin of the resistance change was confirmed by rapid relaxation observed in oscillating magnetic fields at the resonance frequencies of the lattice nuclei ${}^{69}$Ga, ${}^{71}$Ga and ${}^{75}$As.

The resistance change $\Delta R_{yy}$ just after applying dc current is shown in Fig.~5.
\begin{figure}
\includegraphics[width=7cm]{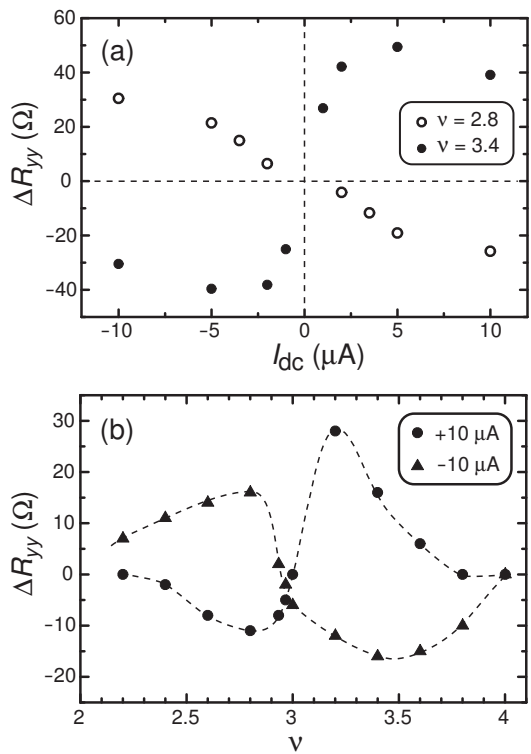}
\caption{
Deviation of $R_{yy}$ from the equilibrium value after applying $I_{\rm dc}$ for $-10~{\rm min} < t < 0~{\rm min}$. $\Delta R_{yy}$ at $t=0$ is shown (a) as a function of $I_{\rm dc}$ for $\nu=2.8$ and 3.4, and (b) as a function of $\nu$ at which $I_{\rm dc}=\pm 10~\mu {\rm A}$ is applied.
Data taken at different cooling cycles are presented in (a) and (b).
}%XXXXX
\end{figure}
We expect a linear relationship $\Delta R_{yy} = \alpha \langle I_z \rangle$ where $\langle I_z \rangle$ is DNP in the vicinity of the right edge.
The absolute value of the coefficient $\alpha$ ($\alpha >0$) is not given since the exact relationship between the edge-bulk coupling and the electron Zeeman energy was not determined.
However, $\Delta R_{yy}$ can be regarded as a semiquantitative measure of DNP induced by $I_{\rm dc}$ since it was measured for the same $\alpha$ at $\nu=3.2$ irrespective of $I_{\rm dc}$ and $\nu$ at which $I_{\rm dc}$ was applied.
As shown in Fig.~5(a), the sign of $\Delta R_{yy}$ depends on the polarity of $I_{\rm dc}$.
It was also found to be reversed with the reversal of the magnetic field.
These observations demonstrate that it is determined by the direction of the Hall electric field for each $\nu$.
In the configuration shown in Fig.~2, the sign of the Hall electric field is positive (plus $x$ direction) for $I >0$.
In Fig.~5(b), $\Delta R_{yy}$ for $I_{\rm dc} = \pm 10~\mu {\rm A}$ is plotted as a function of $\nu$.
It crosses zero almost at $\nu = 3$ for both $I_{\rm dc}$.
The results indicate that the directions of the spin current and the Hall electric field are the same for $\nu > 3$, but are opposite for $\nu<3$.

In Ref.~[\onlinecite{Akera2005}], Akera and Suzuura have shown theoretically that the electron temperature gradient perpendicular to the electric current exhibits quantum oscillations as a function of the position of $\mu$ within the LL structure where the spin splitting is neglected.
It crosses zero when $\mu$ lies near the center of the Landau gap.
In our system, the Zeeman energy spitting ($\sim 2~{\rm K}$), which is much smaller than the cyclotron energy ($\sim 100~{\rm K}$), is comparable to the lattice temperature of 1.6~K \cite{energy}.
At $\nu=3$, $\mu$ is expected to be at the center of the Zeeman gap between the LL($\downarrow, 1$) and LL($\uparrow, 1$).
The observed polarity of the spin current, which is opposite to the heat flow around $\nu=3$, is consistent with the calculation of the electron temperature gradient in Ref.~[\onlinecite{Akera2005}].

Qualitatively, the $\nu$-dependent polarity of the spin current may be explained by a simple picture.
The dissipative electric current parallel to the electric field is carried by electrons in the LL($\downarrow, 1$) and holes in the LL($\uparrow, 1$) around $\nu=3$.
The electric field is almost perpendicular to the channel direction since the Hall resistivity is much larger than the longitudinal resistivity in strong magnetic fields.
Equipotential lines, along which the nondissipative current by the drift motion of electrons flows, are slightly tilted from the channel direction.
Electrons in the LL($\downarrow, 1$) cause the spin current having the same direction as the electric field, which is opposite to the dissipative motion of electrons.
On the other hand, the direction of the spin current carried by holes in the LL($\uparrow, 1$) is opposite to the electric field.
The majority carriers, which determine the polarity of the total spin current, are electrons in the LL($\downarrow, 1$) for $\nu >3$ but are holes in the LL($\uparrow, 1$) for $\nu <3$.
For the consideration of the heat flow, the Landau level energy of the majority carriers measured from $\mu$ is important \cite{Akera2005}.
The polarity of the heat flow is expected to change when $\mu$ crosses the LLs at $\nu \approx 5/2$ and $7/2$.
On the other hand, the direction of the spin current is determined only by the spin of the majority carriers.
In fact, the observed polarity of $\Delta R_{yy}$ does not change in the range $2<\nu<3$ or $3<\nu<4$.

It is worthwhile to discuss the possible spin current and DNP around $\nu=1$ although the experiments were not performed because a very high magnetic field was needed for the sample used.
It is expected that the spin current is carried by skyrmions \cite{Sondhi1993} for $\nu>1$ but by antiskyrmions for $\nu<1$.
Its polarity for $\nu>1$ ($\nu<1$) is the same as that around $\nu=3$ for $\nu>3$ ($\nu<3$).
However, the induced DNP may be very small since the nuclear relaxation rate is considered to be significantly enhanced in a Skyrme crystal state \cite{Hashimoto2002,Tycko1995,Cote1997}.

In summary, we have studied the spin current perpendicular to the electric current in the region of $2<\nu<4$ in a GaAs/AlGaAs 2DES.
A specially designed Hall bar was used in order to detect dynamic nuclear polarization in the vicinity of one of the edges after applying a large electric current.
The observed polarity of DNP, which depends on those of the electric current and magnetic field, indicates that the directions of the spin current and the Hall electric field are the same for $\nu>3$, but are opposite for $\nu<3$.
It is suggested that the spin of the majority carriers of the dissipative current determines the direction of the spin current.

The authors thank H. Akera for helpful discussions.
This work was partly supported by Sumitomo Foundation, Grant-in-Aid for Scientific Research (B) (No. 18340080) and Grant-in-Aid for Scientific Research on Priority Area "Physics of new quantum phases in superclean materials" (No. 18043008) from MEXT, Japan.

\end{document}